%% file: main.tex
 \pdfoutput=1
\documentclass[11pt,a4paper]{article}
\usepackage[hyperref]{emnlp-ijcnlp-2019}
\usepackage{times}
\usepackage{latexsym}
\usepackage{graphicx}
\usepackage{url}
\usepackage{algorithmic}
\usepackage{algorithm}
\usepackage{devanagari}
\usepackage{amsmath}
\usepackage{multicol}
\usepackage[normalem]{ulem}

\aclfinalcopy 

\title{Tale of tails using rule augmented sequence labeling for event extraction}

\author{Ayush Maheshwari, Hrishikesh Patel, Nandan Rathod, \\ \textbf{Ritesh Kumar, Ganesh Ramakrishnan and Pushpak Bhattacharyya} \\
IIT Bombay, Mumbai, India \\
  {\tt \{ayusham,hrishikesh,nandanr,riteshkumar,ganesh,pb\}@cse.iitb.ac.in} \\}

\date{}

\begin{document}
\maketitle
\begin{abstract}

The problem of event extraction is a relatively difficult task for low resource languages due to the non-availability of sufficient annotated data. Moreover, the task becomes complex for tail (rarely occurring) labels wherein extremely less data is available. In this paper, we present a new dataset (InDEE-2019) \footnote{Indic Disaster Event Extraction-2019}
in the disaster domain for multiple Indic languages, collected from news websites. Using this dataset, we evaluate several rule-based mechanisms to augment deep learning based models. We formulate our problem of event extraction as a sequence labeling task and perform extensive experiments to study and understand the effectiveness of different approaches. We further show that tail labels can be easily incorporated by creating new rules without the requirement of large annotated data. 
\end{abstract}
\input{intro}

\input{approach}

\input{experiments}

\bibliography{ref}
\bibliographystyle{acl_natbib}
\end{document}

%% file: intro.tex
\section{Introduction}
Event occurrences involve several entities such as \textit{time}, \textit{date}, \textit{place}, \textit{reason}, \textit{etc}. Event extraction recovers structured representations from the text, often characterised by complex argument and nested events, involving several entities. Entities have associated properties and attributes such as \textit{reason} of the happening, \textit{after-effects} of the events, \textit{etc.} Event extraction has several applications in tasks that include text summarization, knowledge-base construction, machine translation, \textit{etc}. \par
Entity extraction is a pre-requisite for building any event extractor. Additionally, event extraction (EE) often requires the discovery of relations between entities and dependencies between relations from the text. Typically, EE systems find \textit{triggers} and their associated \textit{arguments}. The discovery of \textit{triggers} and \textit{arguments} poses several challenges. Firstly, event detection is highly contextual driven. The same event may appear with different trigger words and a trigger word can evoke different event expressions. For example, consider a sentence
\par `On 29 December 2017 a massive \textbf{fire} broke in Kamala Mills, Mumbai the capital of Maharashtra, \textbf{killed} at least \textbf{14 people} and injured several'. 
\par The trigger word \textit{killed} may co-occur with \textit{fire\_accident} or \textit{attack} or any other keyword associated with an incident involving such a trigger word. Similarly, the trigger word \textit{killed} can evoke different events depending upon the context. Secondly, the presence of a large number of entities requires large annotated data focused on deep learning based systems. Typical deep learning systems require a sufficiently large amount of data for training~\cite{Sun2017RevisitingUE}. Named entity recognition (NER) systems are not effective for extracting a large number of tags given less data. Third, of particular concerns are the labels with very few instances in the dataset, hereafter referred to as `tail labels'. Most event extraction methods are not designed to be fair to tail labels and land up being biased by the more frequent labels. Additionally, the absence of embeddings that capture language models effectively is especially pronounced for low resource languages. This can lead to propagation of errors resulting from the improper initialisation of embeddings for out-of-vocabulary words. \par 
In this paper, we address these challenges to create an end-to-end EE system in the disaster domain in 5 languages, namely, Marathi, English, Hindi, Bengali and
Tamil. We introduce rule augmented deep learning methods and demonstrate the effectiveness of our approach with extensive experiments. We model EE as a sequence labelling task and investigate ways of enhancing state-of-the-art entity extractors by augmenting using rule-based approaches. Our contributions are summarised as follows :
\begin{itemize}
    \item We re-purpose existing rule augmented deep learning models for learning event structures that captures event arguments and their inter-dependencies on disaster domain in low-resource languages.
    \item We conduct extensive experiments on our dataset and demonstrate the importance of rule-augmented deep learning models in improving performance on tail labels.
    \item We release a new dataset named InDEE-2019 that consists of tagged event extraction data in the disaster domain covering five languages: Marathi, English, Hindi, Bengali and Tamil. 
\end{itemize}

\section{Related Work}
Event extraction is a well studied problem, albeit mostly in the English language. Presently, most of the event extraction method consist of deep learning model which requires large annotated training datasets. These data hungry methods work in resources rich language like English \cite{Nguyen2016JointEE} but it fails in low resource languages like Indic languages. Few previous works have used rule based \cite{valenzuela2015domain}, however, the model is not able to learn complex functions of data. Other work \cite{Reschke2014EventEU} uses sources of external information from the knowledge bases to improve the performance of linear chain Conditional random field (CRF) baselines. Therefore, we need hybrid approach to leverage benefits of both methods. Snorkel~\cite{Ratner2017SnorkelRT} deals with candidate extractors and use weak supervision sources to classify the correct set of arguments. However, snorkel requires candidate to be NER tagged and it works well with lesser number of labels. To address this issue, we have augmented rules in different ways with deep learning architectures to learn contextual information along with rules.

%% file: approach.tex
\section{Event Extraction Task}
We focus on the EE task and model as a sequence labeling problem. 
We use the following terminologies throughout the paper:
\begin{itemize}
    \item \textbf{Event Trigger} : The main word that identifies occurrence of the event mentioned in the document.
    \item \textbf{Event Arguments} : Several words that define an events such as \textit{ place , time , reason , after-effects , participant , casualties}.
    
\end{itemize}
We are interested in extracting event triggers, event arguments from the document. Table \ref{data_table} shows the number of labels for each language.

\begin{figure*}[!ht]
    \centering
    \includegraphics[height=4cm,width=0.9\linewidth, keepaspectratio]{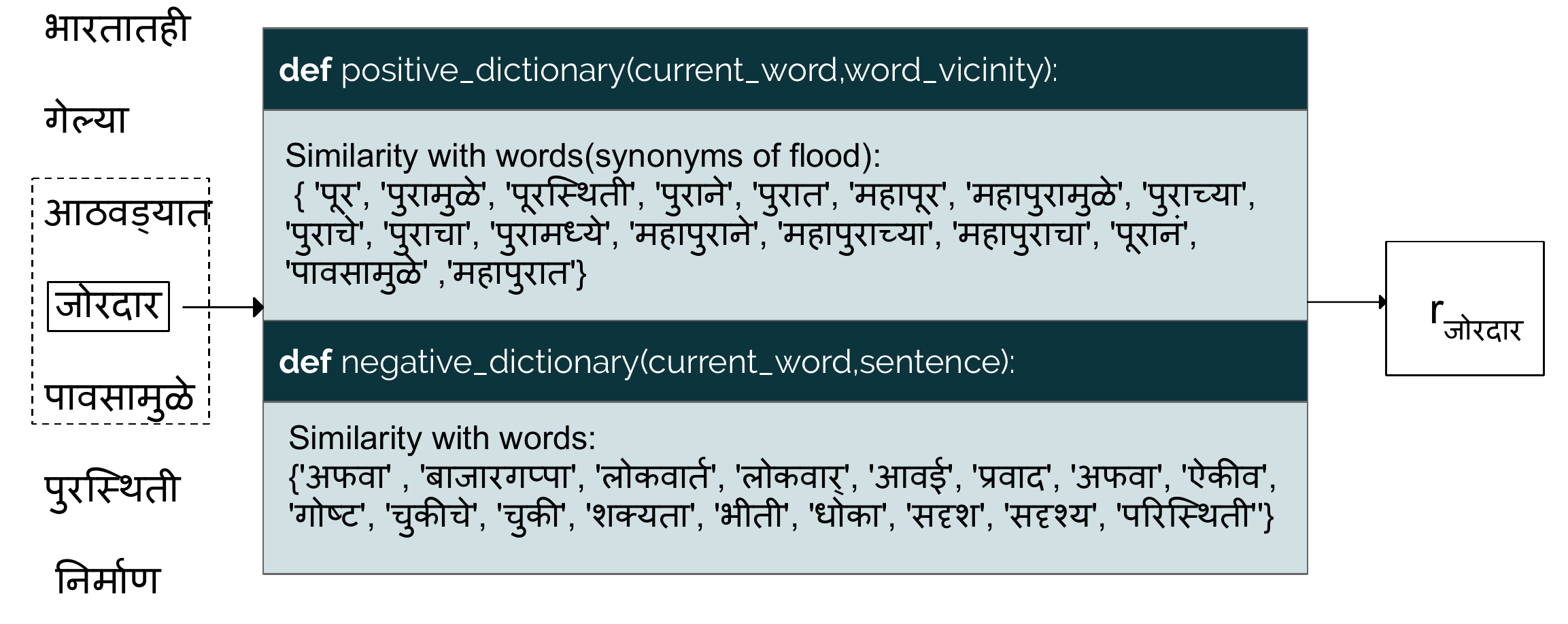}
    \caption{The current word is checked for the similarity with positive and the negative dictionary {in marathi} along with the vicinity words as shown in dotted box.}
    \label{Marathi_Dictionaries}
\end{figure*}

\section{Approach}
In this section, we describe our approach for the event extraction task. We model the extraction as a sequence labeling problem. We adopt Bi-directional LSTM (Long Short Term Memory) and CRF (Conditional Random Fields) \cite{Huang2015BidirectionalLM} for sequence labeling as a baseline and augment it by incorporating information from rules.

\subsection{Bi-directional LSTM}
Recurrent Neural Networks (RNN) has achieved significant gains in sequence labeling tasks. Due to its sequential architecture, RNNs are able to capture previous sequential inputs to predict the output. LSTM are variants of RNN and has outperformed RNNs to capture long-range dependencies in the sequence \cite{Hochreiter1997LongSM}. Recently, state-of-the-art named entity extractors \cite{Lample2016NeuralAF} have used bi-directional LSTMs with CRF. CRF effectively capture transition states during inference, however, they do not perform well on skewed distribution of tags . CRF layer enables to add constraints during inference such that invalid label sequences are discarded from the search space. However, this benefit diminishes on less training data when transition probabilities are not learned effectively.\par
Given an input sequence words $w_1, w_2, \ldots w_n$, LSTM capture current state $\alpha_i$ based on current input word $w_i$ and previous hidden state $\alpha_{i-1}$. Single directional LSTM are not sufficient to capture dependencies from future words. In order to address this issue, a LSTM in reverse direction is used to generate another hidden vector representation $\beta_{i-1}$.\par
It is well documented that LSTMs reduces vanishing gradient problem to model inter-dependencies in long sequences. LSTM output tag probability distribution for each individual word in an input sequence while CRF gives a score for the tag sequence. Due to skewed tag distribution in our dataset, CRF gives very low probabilities for the rarely occurring tags. Therefore, we chose Bi-LSTM as our baseline model and eliminate CRF due to highly skewed tag distribution in our dataset.

\subsection{Rule Definition}
Our approach extends Bi-LSTM with a rule-based approach that generates a rule vector for each word. The rule vectors are built using event anchors to capture class information. We create several dictionaries that correspond to characteristics of event triggers. These dictionaries augment the features learned by the Bi-LSTM. Figure \ref{Marathi_Dictionaries} gives an example of dictionary for developing rules.
\subsubsection{Synonym-based dictionary}
The dictionary corresponds to synonyms of the \textit{trigger} word. We create sets of words that correspond to the synonyms of the trigger word for each language. For instance, synonyms of \textit{flood} are shown in Figure \ref{Marathi_Dictionaries}. 
{Once we create robust dictionary for one language is easy to extend it to other languages. We will refer synonym dictionaries as positive dictionaries.}

\subsubsection{Negative Dictionary}
In several documents, many sentences refer to events that happened in the past {or have a chance of happening}. Such events needs to be ignored by our system {such} that  {\textit{unreasonable mentions of events}} are not considered as \textit{actual} events. For example, ' \textit{There exists a strong \textbf{possibility} of spreading of Malaria after 2015 floods in Mumbai} '. The \textbf{possibility}  word in the sentence , causes the event to classify as \textit{probable event}. Therefore, it is tagged as a negative mention in the annotation. In order to capture these negative instances, we created a negative dictionary as shown in Figure \ref{Marathi_Dictionaries} that contains such words. 

\subsection{Rule Vector}
Given a sentence $s$, containing word sequences $w_1, w_2, \ldots , w_n$, we create rule vector $r_i$ for each word $w_i$. Due to overlapping labels in our tag set, $r_i$ is a multi-hot vector. For instance, any \textit{attack} can belong to both \textit{Normal\_bombing} and \textit{Terrorist\_attack}. 
\par Algorithm \ref{algo_rule} generates rule vector $r_i$ for each word $w_i$ in the sentence $s$. Dimension of $r_i$ is equal to number of labels + 1 (for $other$ tag). If any word $w_i$ in the sentence $s$ found in the negative dictionary, $neg$ then the word should be tagged as $other$.  If any word in window of  \{$w_i-l$, $w_i+l$\} for current word $w_i$, present in particular dictionary $syn_t$ then $w_i$
should be tagged as tag $t$. If none of the word from the window matched with any of the synonym dictionaries then current word $w_i$ should be tagged as $other$. The algorithm for rule vector formation is shown in Algorithm \ref{algo_rule}. We have taken multiple values of $l$ for experimentation. We also used similarity based matching in instead of exact word matching.
\begin{algorithm}
\caption{Algorithm for creating rule vector}\label{algo_rule}
\begin{algorithmic}[1]
\REQUIRE{Sentence $s$ with words $w_1, w_2, \ldots, w_n$, current word $w_i$ in $s$, synonym dictionary $syn_t$ for each label $t$ and negative dictionary $neg$, $2l$ is the window size for $w_i$}, $flag$ indicates word appeared in $syn$ or not.
\ENSURE{ Rule vector $r_i$ for each word $w_i$}
\STATE Initialize each $r_i = [0]$ \# $|r_i|$=number of labels + 1
\STATE Initialize $flag = False$
    \IF {\{$w_1, w_2, \ldots, w_n$\} in $\{neg\}$}
         \STATE $r_i[other]=1$
         \STATE return $r_i$
    \ENDIF
\FOR{each tag $t$}
\IF { $\exists w \in $ \{$w_i-l$, $w_i+l$\} exists in $\{syn_t\}$ }  
         \STATE set  $r_i[t]=1$ \#for corresponding  $tag$ position $t$.
         \STATE $flag= True$
    \ENDIF
\ENDFOR
\IF{ $flag$ is $False$} 
\STATE $r_i[other]= 1$
\ENDIF
\STATE return $r_i$
\end{algorithmic}
\end{algorithm}

\begin{figure*}[!ht]
    \centering
    \includegraphics[width=\linewidth]{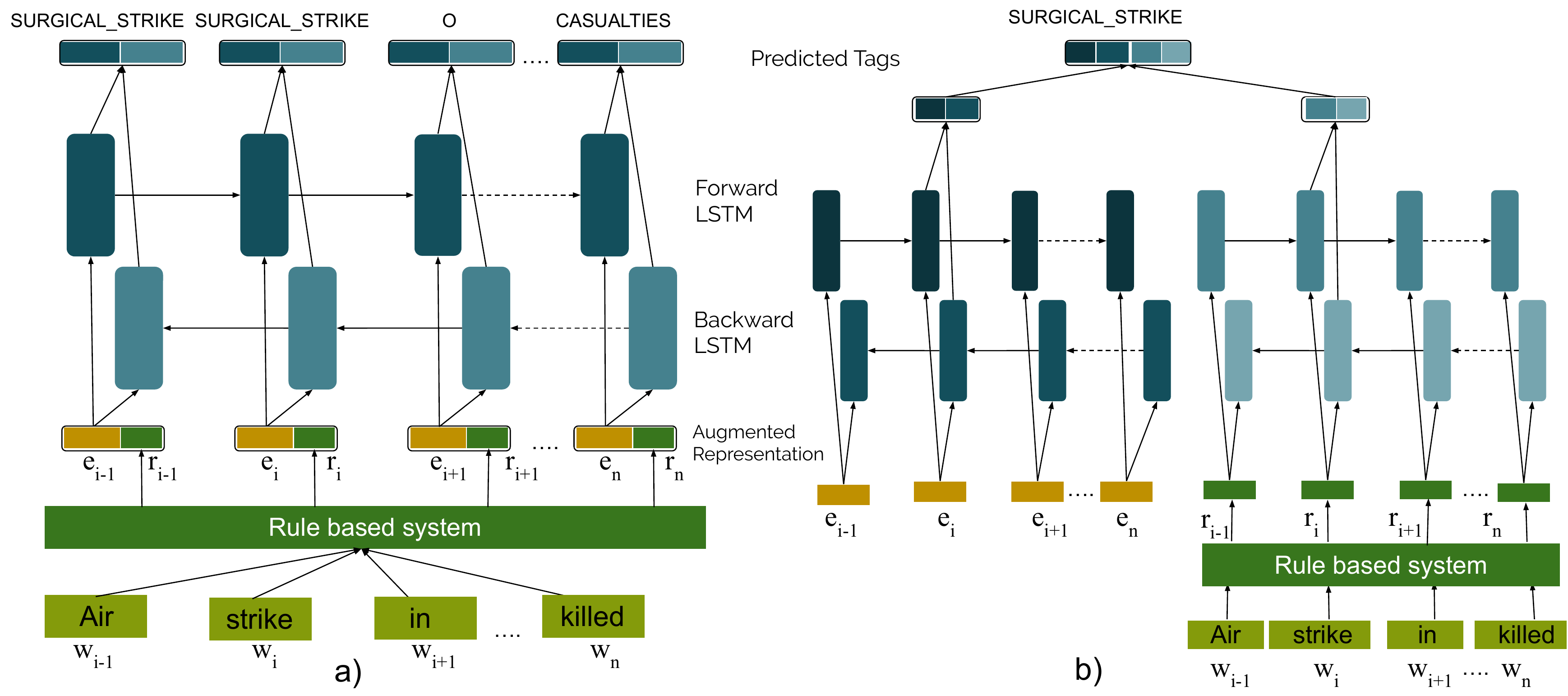}
    \caption{a) Architecture for rule augmentation concatenating with word embeddings. Words sequences are fed to the rule-based system and resultant rule vector is concatenated with respective word vectors and given as input to the Bi-LSTM layer. The hidden vectors are used to retrieve predicted labels. b) Rule vector and word embeddings are given to separate Bi-LSTM and their hidden vectors are concatenated to predict labels. }
    \label{rule_arch}
\end{figure*}
\subsection{Techniques for incorporating rule vector}
We use rules to augment our deep learning labeling architecture using following three methods. The complete architecture is shown in Figure \ref{rule_arch}.
\subsubsection{Augment embeddings with rule vector}\label{B}
In this method, we append the rule vector along with the word representation. In the encoding phase, we extract word representation $e_i$ for word $w_i$ using pre-trained fastText embeddings\cite{fasttext} fine-tuned on our dataset. The word embeddings are appended with the rule vectors $r_i$. The concatenated vector on application of dropouts \cite{Srivastava2014DropoutAS} is fed to the Bi-LSTM that extract essential features and learn representation. 
 
\subsubsection{Explicit rule addition}\label{C}
In addition to the word embeddings fed to the Bi-LSTM, a rule vector through a separate Bi-LSTM is passed. Two parallel Bi-LSTM are fed with word embeddings and rule vectors. The hidden layer representation are concatenated to learn a joint representation of words and rules. The architecture is shown in Figure \ref{rule_arch}.\par
We expect this method to 
retain essential information from rules. In comparison to the concatenation approach proposed in the previous section, we expect it to retain more relevant information by passing rules explicitly.
\begin{table*}[!ht]
\centering
\begin{tabular}{|c|c|c|c|c|c|c|c|c|c|c|}
\hline
Languages & \multicolumn{2}{c|}{Marathi(Mr)} & \multicolumn{2}{c|}{Hindi(Hi)} & \multicolumn{2}{c|}{English(En)} & \multicolumn{2}{c|}{Tamil(Ta)} & \multicolumn{2}{c|}{Bengali(Bn)} \\ \hline
          & Doc            & Sen             & Doc           & Sen            & Doc            & Sen             & Doc           & Sen            & Doc            & Sen             \\ \hline
Train     & 815            & 15920           & 678           & 13184          & 456            & 5378            & 1085          & 15302          & 699            & 18533           \\ \hline
Val       & 117            & 2125            & 150           & 2775           & 56             & 642             & 155           & 2199           & 100            & 2621            \\ \hline
Test      & 233            & 4411            & 194           & 3790           & 131            & 1649            & 311           & 4326           & 199            & 4661            \\ \hline
\#Labels    & \multicolumn{2}{c|}{43}          & \multicolumn{2}{c|}{44}        & \multicolumn{2}{c|}{48}          & \multicolumn{2}{c|}{47}        & \multicolumn{2}{c|}{46}          \\ \hline
\end{tabular}
\caption{InDEE-2019 dataset for five languages, namely, Marathi, Hindi, English, Tamil and Bengali. Number of tags or labels for each dataset and their respective train, validation and test split used in the experiments.}
\label{data_table}
\end{table*}

\subsubsection{Rule projection using distillation}\label{D}
In this method, the knowledge of rules is distilled within deep learning network. This is achieved by biasing the weights of neural networks according to the rule vector. \cite{Hu2016HarnessingDN} proposed a student and teacher model such that model weights are learnt within constraints of rule based system. The teacher-student models bias each other such that weights are learnt by modelling logical rules as constraints and projecting rules within constrained space. Using similar framework we made small changes to make it work to with side information provided by the rules. As given by the eq. \eqref{eq:1}
\begin{equation}
q^{*}(T|w_i) \propto p_{\theta}(T|w_i) \exp{(-C(1-r_i))} \label{eq:1}
\end{equation}
The teacher distribution, $q^{*}(T|w_i)$ is the modified form of the student distribution, $p(T,w_i)$ to fit to transform according to rule vector $r_i$ keeping the basic intuition same, where $T$ is the tag distribution. The formulation modifies the prediction probabilities to have a bias towards the rule vector. For example, $r_i$ being a multi-hot vector can be interpreted as set of labels(tags) assigned by the rules. The formula tries to the bias the student distribution $p_{\theta}(T|w_i)$ by keeping the probability of the tag predicted by the rules, and penalises the probabilities of other labels by a factor of $\exp{(C)}$, where $C$ is an arbitrary constant, we set $C = 1$ for our experiments .
\par Using eq.\eqref{eq:2}, the student tries to mimic the teacher prediction $s^{(t)}_n$  with an imitation parameter $\pi$ and also tries to optimize for ground truth $T_n$. For our experiments we set our imitation parameter to value $0.4$.
\[\theta^{(t+1)} =  \underset{\theta \in \Theta}{\arg\min}\frac{1}{N} \underset{n=1}{\overset{N}{\sum}} (1-\pi)l(T_n,\sigma_{\theta}(w_n))\]
\begin{equation}
+ \pi KL(s_n^{(t)},\sigma_{\theta}(w_n))\label{eq:2}
\end{equation}
Parameters of the student distribution  $\theta$ is updated using the above update rule, $\sigma_{\theta}(w_n)$ is the student prediction at the n-th word in a sequence where $l$ is application specific loss (cross-entropy here), $KL$ is KL divergence loss. Final inference is made using the projected distribution, i.e. the teacher distribution.

\subsection{Word Embeddings}
Due to the very nature of low resource languages, out-of-vocabulary words are common occurrences. To handle such words, we use pre-trained fastText word embeddings~\cite{fasttext} and fine-tune on our corpus. We observe that word coverage of fast-text embeddings on Marathi, Tamil and Bengali is around 50\%. Here, we focused only on building an EE system instead of developing word embeddings.

%% file: experiments.tex
\section{Experiments}
\subsection{Dataset}
We release the new dataset named InDEE-2019 which consists of tagged event extraction data in disaster domain covering five languages: Marathi, English, Hindi, Bengali and Tamil. Dataset statistics can be found in Table \ref{data_table}.

For each language, we crawled disaster related documents from regional news websites. To annotate this documents we followed IOB (Inside-Other-Beginning) tagging scheme proposed by \cite{ramshaw1999text}. IOB tagging helps in differentiating between starting and ending of adjacent tags. In our disaster related documents, adjacent occurrence of same tags forms a phrase. Therefore, we adopted a simpler scheme that merges B and I together and tag under TO (T:Tag and O:Other) scheme. Table {\ref{tab:2}} represents sample sentence from the tagged dataset where first column contains tokens of the sentence, second column represents doc id corresponding to sentence and third column represents tag of token. TO tagging scheme can be seen in third column.  \par
We hired linguistic experts for each language to annotate the dataset as per tagging guidelines. For each language, we divided dataset into three parts train (70\%), validation (10\%) and test (20\%). The train dataset is used to train the model for event extraction task, validation is used for hyper-parameter tuning and test dataset is used for testing our model. \par
This dataset is quite challenging because of multiple reasons. First, there are very large number of tags and very less training data. Moreover, there are closely related tags that made the annotation task difficult. Secondly, data distribution is skewed, therefore very less training data exists for many tags. Third, our main focus is on low resource Indic languages that makes the task more challenging.
\begin{table}[!ht]
\begin{tabular}{|c c c|}
 \hline
 Token & Doc Id & TAG\\
 \hline
    A & 2 & O\\
    moderate & 2 & O\\ 
    intensity & 2 & O\\
earthquake & 2 & EARTHQUAKE\\
measuring & 2 & O\\
4.7 & 2 & MAGNITUDE-ARG\\
hit&2&O\\
Meghalaya&2&PLACE-ARG\\
on&2&O\\
Monday& 2&TIME-ARG\\ \hline
\end{tabular}
\caption{TO Tagging Scheme }
\label{tab:2}
\end{table}

\begin{figure*}[!ht]
    \centering
    \includegraphics[height=6cm, width=\linewidth, keepaspectratio]{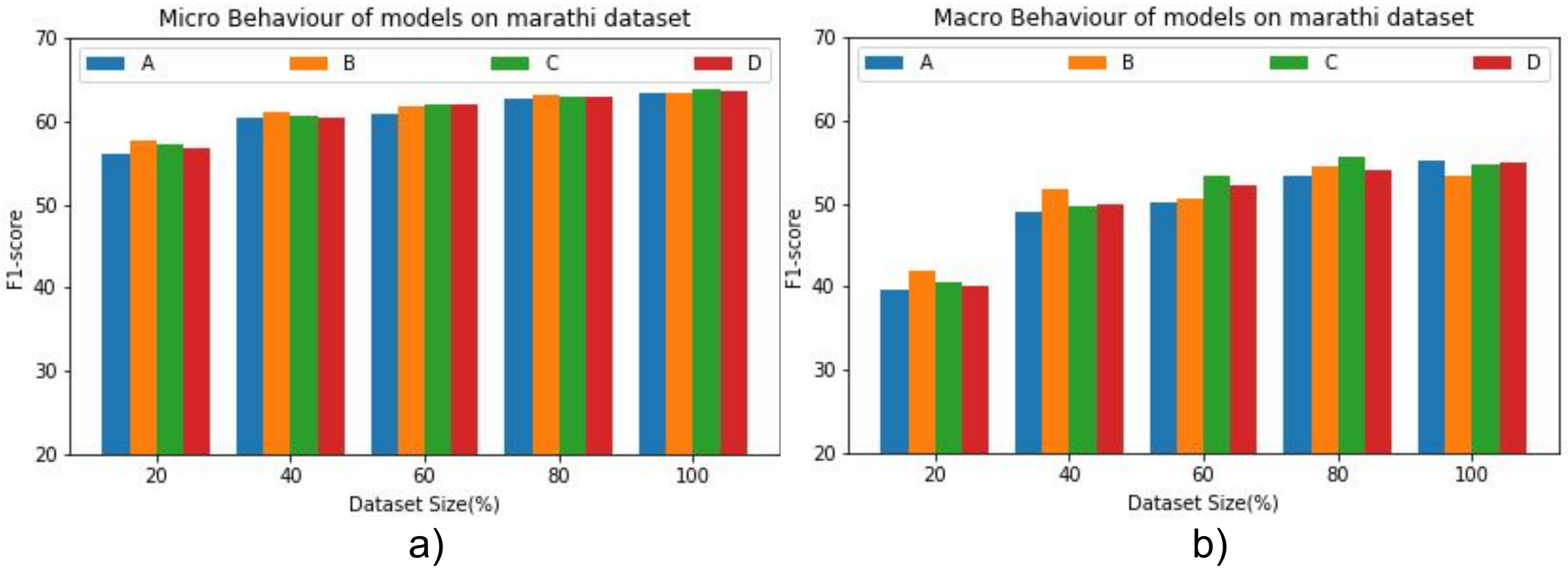}
    \caption{a) Comparison of Micro-F1 scores for different  experiments over various training sizes (in \%) b) Comparison of Macro-F1 scores for different  experiments over various training sizes (in \%)}
    \label{marathi_overall}
\end{figure*}
\subsection{Results and Discussion}

We use Bi-LSTM as our baseline and compare with the proposed three approaches. We conduct experiments on our InDEE dataset on 5 languages, namely, Hindi, Marathi, Bengali, Tamil and English. Our evaluation metric is standard micro-F1 and macro-F1 scores. Micro-F1 score counts the global true positives, false positives and false negatives whereas Macro-F1 captures the average unweighted class scores. Macro does not take class imbalance into consideration. We observe that due to highly skewed label distribution in our dataset, micro score is of more interest to us. For the ease of use, we will call our baseline as A and three proposed approaches as B(\ref{B}), C(\ref{C}), D(\ref{D}).\par
We train our marathi model on 15.9K sentences and predict on 43 labels. Training sets and sizes for all languages are shown in Table \ref{data_table}. We train our model on varying training set size, namely, 20\%, 40\%, 60\%, 80\% and 100\% to ascertain the impact of rules with decreasing amount of dataset.\\
\textbf{Event Extraction} Figure \ref{marathi_overall} shows F1-macro and F1-micro scores for the marathi EE. It is observed that at lesser training instances, our rule based approach is able to outperform baseline on 3K(20\%) and 6K(40\%) training instances. Both macro and micro F1 scores show improvement over baseline model. As training instances increases, deep learning models are able to learn large number of parameters that were difficult to learn on lesser training instances.\par

\begin{figure*}[h]
    \centering
    \includegraphics[height=6cm, width=\linewidth, keepaspectratio]{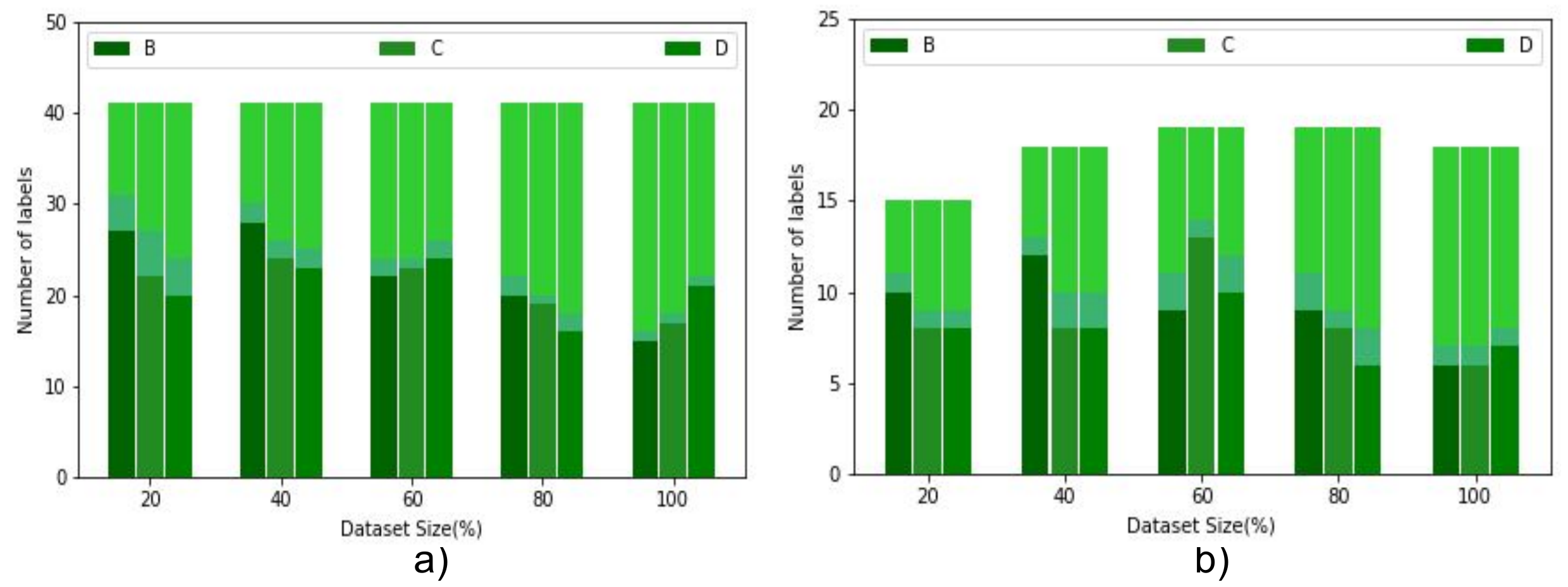}
    \caption{a) Comparison of proposed rule based approaches on improvement over all labels. b) Comparison of proposed approaches over tail labels. Lowest stack represents number of labels shown improvement over baseline, middle stack represent count of labels that has equal score with baseline and upper stack represent count of labels that has lesser score than baseline.}
    \label{marathi_tail_overall}
\end{figure*}

\begin{table}[!ht]

\centering
{%
\begin{tabular}{|c|c|c|c|c|c|c|}
\hline
Language & \multicolumn{3}{c|}{20\%} & \multicolumn{3}{c|}{40\%} \\ \hline
 & B & C & D & B & C & D \\ \hline
Mr & \textbf{10} & 8 & 8 & \textbf{12} & 8 & 8 \\ \hline
Hi & 9 & \textbf{12} & 8 & \textbf{10} & 6 & 6 \\ \hline
En & \textbf{5} & \textbf{5} & 4 & \textbf{5} & 2 & 3 \\ \hline
Ta & \textbf{4} & 3 & \textbf{4} & \textbf{9} & 5 & 6 \\ \hline
Bn & \textbf{8} & 6 & 6 & \textbf{8} & 7 & 3 \\ \hline
\end{tabular}%
}
\caption{Number of tail labels improved on 20\% and 40\% training instances.}
\label{tail_tags_improvement}
\end{table}

\begin{table}[!ht]
\centering
\begin{tabular}{|c|c|c|c|c|c|c|}
\hline
Language & \multicolumn{3}{c|}{20\%} & \multicolumn{3}{c|}{40\%} \\ \hline
 & B & C & D & B & C & D \\ \hline
Mr & \textbf{27} & 22 & 20 & \textbf{28} & 24 & 23 \\ \hline
Hi & \textbf{19} & \textbf{19} & 16 & \textbf{21} & 16 & 17 \\ \hline
En & 20 & \textbf{21} & 20 & \textbf{16} & 9 & 9 \\ \hline
Ta & \textbf{13} & 10 & 11 & \textbf{16} & 14 & 12 \\ \hline
Bn & \textbf{14} & 12 & 11 & \textbf{14} & 11 & 9 \\ \hline
\end{tabular}
\caption{Number of total labels improving over baseline on 5 languages. We have shown results for 20\% and 40\% training instances.}
\label{tails_overall_improvement}
\end{table}
Table \ref{tab:4} shows macro-F1 and micro-F1 scores for all 5 languages. We observe that on smaller training instances, implicit rule approach tend to perform better on Marathi \& Hindi. Most of the cases shows that our methods performs better than the baseline on less training data. Due to better word representation and data annotation on English, the parameters are even learnt on smaller dataset by Bi-LSTM. However, for low resource languages, Bi-LSTM is not able to learn parameters on similar training instances.\\
\textbf{Tail Labels} Most deep learning based methods are not able to capture tail labels due to lesser training data. However, this is of interest to us since most real world data has small size and has large label set.  Figure \ref{marathi_tail_overall} shows improvement of F1-score for each classes over baselines. We only considered scores that are greater than baseline scores. We observe that more classes are improved by including rules over baselines. At lesser training instances, we observe that large number of tags are correctly classified. Table \ref{tails_overall_improvement} shows number of tags improved over baseline. In order to further prove effectiveness of rule based approach, we tested the improvements of classes over tail labels. We chose those tails labels whose sum forms 5\% of total training set instances. We notice significant improvement of tags over 20\% and 40\% training instances. We can see the number of tail labels improved over baseline in table \ref{tail_tags_improvement}. Moreover, we calculated micro-F1 scores for the tails labels. The results are shown in Table \ref{tab:3}. We notice significant improvement on tails for 20\% and 40\% training instances. 
It is clear from the results that at lesser training instances, our approach is effective in capturing tail labels. While working with low resource languages and lesser annotated data, combination of rules and Bi-LSTM gives higher score. Additionally, rule based approach is effective at capturing tail labels. 
\begin{table}[!ht]
\centering
\begin{tabular}{|c|c|c|c|c|c|c|}
\hline
\textbf{L}                  & \textbf{M}             & \textbf{20}                                                   & \textbf{40}                            & \textbf{60}                            & \textbf{80}                            & \textbf{100}                           \\ \hline
                                   & \textbf{A}            & 39.76                                                         & 42.67                                  & 46.69                                  & 52.85                                  & 53.88                                  \\ \cline{2-7} 
                                   & \textbf{B}    & {\color[HTML]{333333} \textbf{44.62}} & \textbf{47.44} & 48.91                                  & 54.3                                   & 50.71                                  \\ \cline{2-7} 
                                   & \textbf{C} & 42.45                                                         & 43.02                                  & \textbf{53.05} & \textbf{55.54} & \textbf{53.77} \\ \cline{2-7} 
{\textbf{Mr}} & \textbf{D}       & 41.72                                                         & 43.06                                  & 48.83                                  & 50.93                                  & 51.25                                  \\ \hline
                                   & \textbf{A}            & 29.57                                                         & 42.68                                  & 48.11                                  & 47.35                                  & 46.99                                  \\ \cline{2-7} 
                                   & \textbf{B}    & 31.33                                                         & \textbf{46.35} & 47.15                                  & \textbf{48.65} & 46.52                                  \\ \cline{2-7} 
                                   & \textbf{C} & \textbf{34.08}                        & 42.87                                  & \textbf{49.21} & 46.35                                  & \textbf{47.29} \\ \cline{2-7} 
{\textbf{Hi}}   & \textbf{D}       & 29.87                                                         & 42.84                                  & 47.58                                  & 47.87                                  & 46.45                                  \\ \hline
                                   & \textbf{A}            & 56.96                                                         & \textbf{64.58} & 73.7                                   & 73.92                                  & 82.29                                  \\ \cline{2-7} 
                                   & \textbf{B}    & 61.65                                                         & 63.66                                  & \textbf{75.34} & \textbf{75.31} & 82.71                                  \\ \cline{2-7} 
                                   & \textbf{C} & 58.51                                                         & 63.78                                  & 74.47                                  & 73.64                                  & \textbf{83.25} \\ \cline{2-7} 
{\textbf{En}} & \textbf{D}       & \textbf{63.85}                        & 62.24                                  & 71.67                                  & 69.06                                  & 79.01                                  \\ \hline
                                   & \textbf{A}            & 44.88                                                         & 55.94                                  & \textbf{59.64} & 62.15                                  & 62.62                                  \\ \cline{2-7} 
                                   & \textbf{B}    & \textbf{44.95}                        & \textbf{57.55} & 58.37                                  & 62.31                                  & 60.55                                  \\ \cline{2-7} 
                                   & \textbf{C} & 44.36                                                         & 55.75                                  & 57.06                                  & \textbf{64.08} & 62.73                                  \\ \cline{2-7} 
{\textbf{Ta}}   & \textbf{D}       & 41.59                                                         & 50.93                                  & 55.34                                  & 60.75                                  & \textbf{63.37} \\ \hline
                                   & \textbf{A}            & 38.56                                                         & 49.69                                  & 42.06                                  & 44.29                                  & \textbf{49.38} \\ \cline{2-7} 
                                   & \textbf{B}    & \textbf{42.92}                        & \textbf{51.38} & 38.26                                  & 47.57                                  & 48.3                                   \\ \cline{2-7} 
                                   & \textbf{C} & 42.7                                                          & 50.11                                  & \textbf{42.99} & \textbf{47.65} & 42.15                                  \\ \cline{2-7} 
{\textbf{Bn}} & \textbf{D}       & 41.41                                                         & 46.54                                  & 39.34                                  & 42.33                                  & 44.72                                  \\ \hline
\end{tabular}
\caption{Comparison of Micro F1-score consisting tail labels for all 5 languages. L: Language, M: Models}
\label{tab:3}
\end{table}

\begin{table*}[!h]
\begin{tabular}{|c|c|c|c|c|c|c|c|c|c|c|c|}
\hline
\textbf{Lang}                 & \textbf{Model} & \multicolumn{2}{c|}{\textbf{20\%}}                                                                       & \multicolumn{2}{c|}{\textbf{40\%}}                                                & \multicolumn{2}{c|}{\textbf{60\%}}                        & \multicolumn{2}{c|}{\textbf{80\%}} & \multicolumn{2}{c|}{\textbf{100\%}} \\ \hline
                              &                & Micro                                                         & Macro                                  & Micro                                  & Macro                                  & Micro                                  & Macro          & Micro           & Macro          & Micro           & Macro           \\ \hline
                              & \textbf{A}     & 56.04                                                         & 39.58                                  & 60.39                                  & 48.98                                  & 60.95                                  & 50.19          & 62.76           & 53.26          & 63.37           & 55.2            \\ \cline{2-12} 
                              & \textbf{B}     & {\color[HTML]{333333} \textbf{57.75}} & \textbf{41.89} & \textbf{61.04}                         & \textbf{51.72}                         & 61.86                                  & 50.66          & \textbf{63.22}  & 54.43          & 63.35           & 53.36           \\ \cline{2-12} 
                              & \textbf{C}     & 57.26                                                         & 40.59                                  & 60.67          & 49.64          & \textbf{62.01} & \textbf{53.42} & 62.8            & \textbf{55.54} & \textbf{63.69}  & 54.77           \\ \cline{2-12} 
{\textbf{Mr}} & \textbf{D}     & 56.68                                                         & 40.14                                  & 60.39                                  & 49.82                                  & 61.88                                  & 52.07          & 62.99           & 54.06          & 63.55           & \textbf{54.97}  \\ \hline
                              & \textbf{A}     & 48.56                                                         & 37.73                                  & 51.63                                  & 42.71                                  & 53.12                                  & 46.08          & 54.09           & \textbf{46.83} & \textbf{56.93}  & \textbf{47.12}  \\ \cline{2-12} 
                              & \textbf{B}     & 48.44                                                         & \textbf{38.54} & \textbf{51.67}                         & \textbf{43.44} & \textbf{54.87}                         & 44.2           & \textbf{55.11}  & 46.18          & 56.01           & 46.33           \\ \cline{2-12} 
                              & \textbf{C}     & \textbf{49.34}                        & 38.29                                  & 51.54          & 41.86                                  & 54.23          & \textbf{47.27} & 53.16           & 45.19          & 55.81           & 46.4            \\ \cline{2-12} 
{\textbf{Hi}} & \textbf{D}     & 48.85                                                         & 38.17                                  & 51.22                                  & 41.3                                   & 52.65                                  & 45.45          & 54.18           & 45.53          & 54.93           & 45.88           \\ \hline
                              & \textbf{A}     & 64.79                                                         & 50.39          & \textbf{76.51}                         & \textbf{69.07}                         & \textbf{81.59}                         & 75.85          & 81.75           & 77.98          & 85.76           & 82.44           \\ \cline{2-12} 
                              & \textbf{B}     & 66.47                                                         & 51.71                                  & 75.94          & 68.81          & 81.44                                  & \textbf{76.73} & \textbf{83.88}  & \textbf{80.01} & \textbf{86.66}  & \textbf{83.44}  \\ \cline{2-12} 
                              & \textbf{C}     & \textbf{66.56}                                                & \textbf{54.82}                         & 76.32                                  & 68.63                                  & 81.46          & 74.91          & 82.11           & 78.36          & 86.48           & 82.69           \\ \cline{2-12} 
{\textbf{En}} & \textbf{D}     & 65.95                                 & 52.54                                  & 75.24                                  & 67.06                                  & 81.4                                   & 73.06          & 82.11           & 74.43          & 86.04           & 81.04           \\ \hline
                              & \textbf{A}     & 67.44                                                         & \textbf{61.19}                         & \textbf{70.99} & 64.01                                  & \textbf{73.97}                         & \textbf{67.19} & 73.62           & \textbf{69.77} & 73.62           & \textbf{69.77}  \\ \cline{2-12} 
                              & \textbf{B}     & \textbf{68.28}                        & 61             & 70.82                                  & 63.53                                  & 72.95                                  & 66.3           & \textbf{74.3}   & 69.1           & \textbf{74.3}   & 69.1            \\ \cline{2-12} 
                              & \textbf{C}     & 67.96                                                         & 59.13                                  & 70.85                                  & \textbf{64.08} & 72.8                                   & 65.84          & 73.7            & 69.33          & 73.7            & 69.33           \\ \cline{2-12} 
{\textbf{Ta}} & \textbf{D}     & 66.25                                                         & 58.42                                  & 70.03                                  & 60.36                                  & 72.78          & 63.84          & 73.69           & 69.2           & 73.69           & 69.2            \\ \hline
                              & \textbf{A}     & \textbf{64.26}                                                & 38.07                                  & \textbf{66.58}                         & \textbf{46.82}                         & \textbf{67.52} & \textbf{43.3}  & 67.78           & 45.23          & \textbf{67.94}  & \textbf{47.94}  \\ \cline{2-12} 
                              & \textbf{B}     & 63.9                                  & \textbf{41.21} & 65.97                                  & 46.61                                  & 67.24                                  & 40.96          & 67.61           & \textbf{46.12} & 67.81           & 45.56           \\ \cline{2-12} 
                              & \textbf{C}     & 63.16                                                         & 40.72                                  & 66.24          & 44.75          & 67.25                                  & 42.54          & \textbf{68}     & 45.23          & 67.71           & 42.15           \\ \cline{2-12} 
{\textbf{Bn}} & \textbf{D}     & 63.28                                                         & 38.16                                  & 65.86                                  & 40.73                                  & 65.81                                  & 37.31          & 66.14           & 40.64          & 66.61           & 43.78           \\ \hline
\end{tabular}
\caption{Overall Micro and Macro F1-score for all languages and different training set sizes}
\label{tab:4}
\end{table*}

\section{Conclusion}
In this paper, we introduce a hybrid approach for automatic extraction of events and arguments. We present a new dataset in the disaster domain for five languages consisting of large number of tags than usual datasets. We propose several variants of rule based system to augment deep learning based models. Extensive experimental results demonstrate that our rule augmented methods outperforms deep learning based models on lesser annotated data and low resource languages. We further shows more improvement on tail labels using our approach. For future work, we plan to integrate cross linking between events and its arguments.

%% file: main.bbl
\begin{thebibliography}{12}
\expandafter\ifx\csname natexlab\endcsname\relax\def\natexlab#1{#1}\fi

\bibitem[{Bojanowski et~al.(2017)Bojanowski, Grave, Joulin, and
  Mikolov}]{fasttext}
Piotr Bojanowski, Edouard Grave, Armand Joulin, and Tomas Mikolov. 2017.
\newblock Enriching word vectors with subword information.
\newblock \emph{Transactions of the Association for Computational Linguistics},
  5:135--146.

\bibitem[{Hochreiter and Schmidhuber(1997)}]{Hochreiter1997LongSM}
Sepp Hochreiter and J{\"u}rgen Schmidhuber. 1997.
\newblock Long short-term memory.
\newblock \emph{Neural Computation}, 9:1735--1780.

\bibitem[{Hu et~al.(2016)Hu, Ma, Liu, Hovy, and Xing}]{Hu2016HarnessingDN}
Zhiting Hu, Xuezhe Ma, Zhengzhong Liu, Eduard~H. Hovy, and Eric~P. Xing. 2016.
\newblock Harnessing deep neural networks with logic rules.
\newblock \emph{CoRR}, abs/1603.06318.

\bibitem[{Huang et~al.(2015)Huang, Xu, and Yu}]{Huang2015BidirectionalLM}
Zhiheng Huang, Wei Xu, and Kai Yu. 2015.
\newblock Bidirectional lstm-crf models for sequence tagging.
\newblock \emph{CoRR}, abs/1508.01991.

\bibitem[{Lample et~al.(2016)Lample, Ballesteros, Subramanian, Kawakami, and
  Dyer}]{Lample2016NeuralAF}
Guillaume Lample, Miguel Ballesteros, Sandeep Subramanian, Kazuya Kawakami, and
  Chris Dyer. 2016.
\newblock Neural architectures for named entity recognition.
\newblock In \emph{HLT-NAACL}.

\bibitem[{Nguyen et~al.(2016)Nguyen, Cho, and Grishman}]{Nguyen2016JointEE}
Thien~Huu Nguyen, Kyunghyun Cho, and Ralph Grishman. 2016.
\newblock Joint event extraction via recurrent neural networks.
\newblock In \emph{HLT-NAACL}.

\bibitem[{Ramshaw and Marcus(1999)}]{ramshaw1999text}
Lance~A Ramshaw and Mitchell~P Marcus. 1999.
\newblock Text chunking using transformation-based learning.
\newblock In \emph{Natural language processing using very large corpora}, pages
  157--176. Springer.

\bibitem[{Ratner et~al.(2017)Ratner, Bach, Ehrenberg, Fries, Wu, and
  R{\'e}}]{Ratner2017SnorkelRT}
Alexander Ratner, Stephen~H. Bach, Henry~R. Ehrenberg, Jason~Alan Fries, Sen
  Wu, and Christopher R{\'e}. 2017.
\newblock Snorkel: Rapid training data creation with weak supervision.
\newblock \emph{Proceedings of the VLDB Endowment. International Conference on
  Very Large Data Bases}, 11 3:269--282.

\bibitem[{Reschke et~al.(2014)Reschke, Jankowiak, Surdeanu, Manning, and
  Jurafsky}]{Reschke2014EventEU}
Kevin Reschke, Martin Jankowiak, Mihai Surdeanu, Christopher~D. Manning, and
  Daniel Jurafsky. 2014.
\newblock Event extraction using distant supervision.
\newblock In \emph{LREC}.

\bibitem[{Srivastava et~al.(2014)Srivastava, Hinton, Krizhevsky, Sutskever, and
  Salakhutdinov}]{Srivastava2014DropoutAS}
Nitish Srivastava, Geoffrey~E. Hinton, Alex Krizhevsky, Ilya Sutskever, and
  Ruslan~R. Salakhutdinov. 2014.
\newblock Dropout: a simple way to prevent neural networks from overfitting.
\newblock \emph{Journal of Machine Learning Research}, 15:1929--1958.

\bibitem[{Sun et~al.(2017)Sun, Shrivastava, Singh, and
  Gupta}]{Sun2017RevisitingUE}
Chen Sun, Abhinav Shrivastava, Saurabh Singh, and Abhinav Gupta. 2017.
\newblock Revisiting unreasonable effectiveness of data in deep learning era.
\newblock \emph{2017 IEEE International Conference on Computer Vision (ICCV)},
  pages 843--852.

\bibitem[{Valenzuela-Esc{\'a}rcega et~al.(2015)Valenzuela-Esc{\'a}rcega,
  Hahn-Powell, Surdeanu, and Hicks}]{valenzuela2015domain}
Marco~A Valenzuela-Esc{\'a}rcega, Gus Hahn-Powell, Mihai Surdeanu, and Thomas
  Hicks. 2015.
\newblock A domain-independent rule-based framework for event extraction.
\newblock \emph{Proceedings of ACL-IJCNLP 2015 System Demonstrations}, pages
  127--132.

\end{thebibliography}
